\documentclass[aps, reprint, amsmath, amssymb, amsfonts, floatfix, intlimits,superscriptaddress]{revtex4-1}
\pdfoutput=1
\usepackage{graphicx}
\usepackage{dcolumn}
\usepackage{bm}
\usepackage{color} 
\usepackage{float}

\definecolor{deepBlue}{RGB}{48, 71, 210}
\definecolor{myGreen}{RGB}{200, 38, 216}
\definecolor{APSBlue}{RGB}{46, 48, 146}
\usepackage{url}
\usepackage{color}
\usepackage[colorlinks = true,
	linkcolor = blue,
	urlcolor  = blue,
	citecolor = blue,
	anchorcolor = blue]{hyperref}
	
\begin{document}

\title{Atomic scale characterization of graphene p-n junctions for electron-optical applications}

\affiliation{
		Department of Physics, Columbia University, New York, New York 10027, United States\looseness=-1}
\affiliation{
		Department of Electrical and Computer Engineering, University of Virginia, Charlottesville, VA 22904, United States\looseness=-1}
\affiliation{
		Department of Physics, University of Virginia, Charlottesville, VA 22904, United States\looseness=-1}
\affiliation{
		College of Nanoscale Science and Engineering, The State University of New York at Albany, Albany, New York 12203, United States\looseness=-1}
\affiliation{
	    IBM T. J. Watson Research Center, Yorktown Height, New York 10598, United States\looseness=-1}
\affiliation{
		Microsystems Technology Laboratories, Massachusetts Institute of Technology, Cambridge, Massachusetts 02139, United States\looseness=-1}
\affiliation{Current address: 
	    Institute for Nanoelectronic Devices and Quantum Computing, Fudan University, Shanghai 200438, P.R.China\looseness=-1}
\affiliation{Current address: Intel Corp., Santa Clara, CA 95054, USA.}
\affiliation{Current address: 
	Department of Materials Science and Engineering, Massachusetts Institute of Technology, Cambridge, Massachusetts 02139, United States\looseness=-1}
\author{Xiaodong Zhou}
	\affiliation{
		Department of Physics, Columbia University, New York, New York 10027, United States\looseness=-1}
	\affiliation{
	    IBM T. J. Watson Research Center, Yorktown Height, New York 10598, United States\looseness=-1}
	\affiliation{Current address: 
	    Institute for Nanoelectronic Devices and Quantum Computing, Fudan University, Shanghai 200438, P.R.China\looseness=-1}
\author{Alexander Kerelsky}
	\affiliation{
		Department of Physics, Columbia University, New York, New York 10027, United States\looseness=-1}
\author{Mirza M. Elahi}
	\affiliation{
		Department of Electrical and Computer Engineering, University of Virginia, Charlottesville, VA 22904, United States\looseness=-1}
\author{Dennis Wang}
	\affiliation{
		Department of Physics, Columbia University, New York, New York 10027, United States\looseness=-1}
\author{K. M. Masum Habib}
	\affiliation{
		Department of Electrical and Computer Engineering, University of Virginia, Charlottesville, VA 22904, United States\looseness=-1}
	\affiliation{Current address: Intel Corp., Santa Clara, CA 95054, USA.}
\author{Redwan N. Sajjad}
	\affiliation{
		Microsystems Technology Laboratories, Massachusetts Institute of Technology, Cambridge, Massachusetts 02139, United States\looseness=-1}
\author{Pratik Agnihotri}
	\affiliation{
		College of Nanoscale Science and Engineering, The State University of New York at Albany, Albany, New York 12203, United States\looseness=-1}
\author{Ji Ung Lee}
	\affiliation{
		College of Nanoscale Science and Engineering, The State University of New York at Albany, Albany, New York 12203, United States\looseness=-1}
\author{Avik W. Ghosh}
	\affiliation{
		Department of Electrical and Computer Engineering, University of Virginia, Charlottesville, VA 22904, United States\looseness=-1}
	\affiliation{
		Department of Physics, University of Virginia, Charlottesville, VA 22904, United States\looseness=-1}
\author{Frances M. Ross}
	\altaffiliation[Correspondence to: ]{
	    \href{mailto:apn2018@columbia.edu}{apn2018@columbia.edu}, \href{mailto:fmross@mit.edu}{fmross@mit.edu}\looseness=-1}
	\affiliation{
	    IBM T. J. Watson Research Center, Yorktown Height, New York 10598, United States\looseness=-1}
	\affiliation{Current address: 
	Department of Materials Science and Engineering, Massachusetts Institute of Technology, Cambridge, Massachusetts 02139, United States\looseness=-1}
\author{Abhay N. Pasupathy}
	\altaffiliation[Correspondence to: ]{
	    \href{mailto:apn2018@columbia.edu}{apn2018@columbia.edu}, \href{mailto:fmross@mit.edu}{fmross@mit.edu}\looseness=-1}
	\affiliation{
		Department of Physics, Columbia University, New York, New York 10027, United States\looseness=-1}

\date{\today}
\begin{abstract}
	Graphene p-n junctions offer a potentially powerful approach towards controlling electron trajectories via collimation and focusing in ballistic solid-state devices. The ability of p-n junctions to control electron trajectories depends crucially on the doping profile and roughness of the junction. Here, we use four-probe scanning tunneling microscopy and spectroscopy (STM/STS) to characterize two state-of-the-art graphene p-n junction geometries at the atomic scale, one with CMOS polySi gates and another with naturally cleaved graphite gates. Using spectroscopic imaging, we characterize the local doping profile across and along the p-n junctions. We find that realistic junctions exhibit non-ideality both in their geometry as well as in the doping profile across the junction. We show that the geometry of the junction can be improved by using the cleaved edge of van der Waals metals such as graphite to define the junction. We quantify the geometric roughness and doping profiles of junctions experimentally and use these parameters in Nonequilibrium Green's Function based simulations of focusing and collimation in these realistic junctions. We find that for realizing Veselago focusing, it is crucial to minimize lateral interface roughness which only natural graphite gates achieve, and to reduce junction width, in which both devices under investigation underperform. We also find that carrier collimation is currently limited by the non-linearity of the doping profile across the junction. Our work provides benchmarks of the current graphene p-n junction quality and provides guidance for future improvements.

	\end{abstract}
\maketitle

\section{Introduction}
\begin{figure*}[t]
	\includegraphics[width=\linewidth]
	{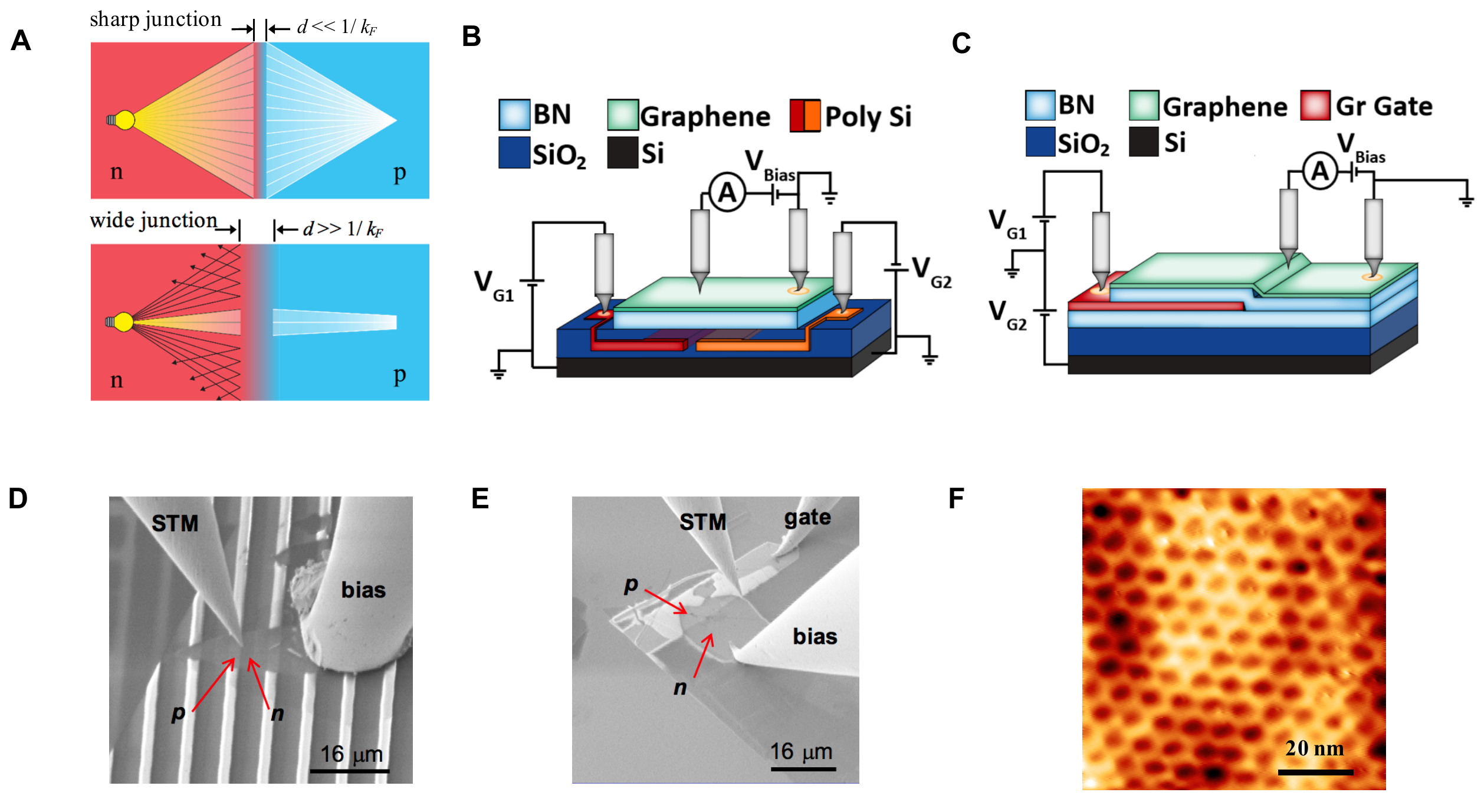}
	\caption{
		\label{fig:fig1}
		(\textbf{a}) Schematic of two electron optical functions for a graphene p-n junction: Veselago lensing for a sharp junction ($k_F d \ll 1$) and collimation for a wide junction ($k_F d \gg 1$). (\textbf{b}) Schematic of the polySi gate device and experimental setup. Two independent local gate $V_{G1}$ and $V_{G2}$ are supplied through a split buried polysilicon to the graphene forming a p-n junction. (\textbf{c}) Schematic of the graphite gate device and experimental setup. A global gate $V_{G2}$ and a partial gate $V_{G1}$ are supplied through SiO$_2$/Si and graphite respectively to the graphene on top. A p-n junction forms at the sharp graphite edge. (\textbf{d-e}) SEM secondary electron images taken under gating conditions for polySi gate device (d) and graphite gate device (e). \textit{p-type} (hole-like) and \textit{n-type} (electron-like) regions display different contrast in terms of secondary electron emission, as denoted by red arrows. Also labeled are the gating probe, bias probe and scanning probe used in the experiment. (\textbf{f}) A representative STM topographic image of graphene on hBN in a polySi gate device. The image shows a moir\'e pattern due to lattice mismatch between graphene and hBN. STM topography set points are 0.6 V, 20 pA.} 
\end{figure*}

    The discovery of graphene as a new two-dimensional electron system with high intrinsic mobility and photon-like band structure has led to a surge of interest in applying it to implement solid-state electron optics systems where the ballistic motion of electrons is explored in order to manipulate the flow of an electron beam \cite{van1995confined}. Ballistic electron transport in GaAs two-dimensional electron gas already allows for carrier steering using electric and magnetic fields \cite{van1989coherent, molenkamp1990electron, spector1990electron, yacoby1994unexpected}. As an example, a local electrostatic potential in GaAs can be used to confine electrons laterally and thus control their flow in a device. However, such performance is limited to cryogenic temperatures in GaAs. Graphene displays ballistic transport over micron length scales at room temperature, renewing interest in solid-state electron optics \cite{dean2010boron, wang2013one, williams2011gate, kim2016valley, rickhaus2015snake, taychatanapat2015conductance}. However, the Dirac dispersion of graphene implies that local electrostatic potentials are not effective at carrier confinement due to Klein tunneling \cite{katsnelson2006chiral}. While electron flow in graphene cannot be turned off by doping, it was realized that the carrier trajectory can be modified by creating electrostatic doping profiles. In particular, two important predictions have been made for electron flow through p-n junctions \cite{cheianov2006selective, cheianov2007focusing, allain2011klein}, in which an interface is created between a \textit{p-type} (hole-like) and an \textit{n-type} (electron-like) region on the same graphene sheet \cite{huard2007transport, ozyilmaz2007electronic}. These predictions are shown schematically in Fig. \ref{fig:fig1}a. The first prediction (upper panel) is for a sharp junction where the junction width $d$ is much smaller than the electron's Fermi wavelength $k_F d \ll 1$. In this case, electrons incident on the p-n interface from a point source will be focused to a point on the other side of the p-n interface \cite{cheianov2007focusing}, a phenomenon termed Veselago lensing. The second prediction (lower panel) is made for a wide junction where $k_F d \gg 1$. In this case, electrons incident on the junction are strongly collimated, with an angle-selective transmission probability given by $T(\theta) \sim e^{-\pi k_F \frac{d}{2} \sin ^2 \theta}$. Therefore, only electrons near normal incidence transmit through the junction while the rest are reflected \cite{cheianov2006selective}. Both focusing and collimation, arising respectively from the plane wave and Bloch parts of the electron wavefunction, can be used to control the electron trajectories, and have given rise to a number of device concepts \cite{vakil2011transformation, sajjad2011high, peterfalvi2012intraband, sajjad2013manipulating, jang2013graphene, park2008electron, zhang2017focusing, tan2017graphene}. Two examples are graphene p-n junction based field effect transistors and radio frequency (RF) switches where multiple angled p-n junctions can be used to turn on and off current flow through the device \cite{sajjad2013manipulating, tan2017graphene}, and enhanced Rudermann-Kittel-Kasuya-Yosida (RKKY) interaction for scalable graphene-based spintronics devices by utilizing Veselago focusing of graphene p-n junctions \cite{zhang2017focusing}. Note that for an RF switch, a modest current on-off ratio much lower than the ballistic 10$^4$ or even 100 would suffice.

    The theoretical predictions of geometric optics-like carrier transport in graphene have led to a number of experimental efforts to realize these predictions, primarily via measuring electron transport across one or more p-n junctions. Lee \textit{et al.} conducted transport measurements across a p-n interface where signatures of lensing were observed, albeit after a background subtraction procedure \cite{lee2015observation}. Chen \textit{et al.} demonstrated negative refraction in a graphene p-n junction, but were able to observe the effect only under a magnetic field \cite{chen2016electron}. Barnard \textit{et al.} recently reported an electron beam collimator in graphene based on collinear pairs of slits \cite{barnard2017absorptive}, but clean collimation has yet to be realized in single p-n junctions \cite{park2008electron, cheianov2006selective}. Sajjad \textit{et al.} proposed a collimation-based field effect transistor using two graphene p-n junctions aligned at different angles and predicted a $\sim$10$^4$ on-off ratio for ballistic trajectories with no edge scattering, $\sim$10$^2$ for perfect edges in a 1 $\mu$m wide device, and $\sim$10$^1$ in the presence of edge roughness \cite{sajjad2013manipulating, jang2013graphene}. Experimentally, a $R_{off}/R_{on} \sim$ 1.3 was reported before in a similar device \cite{morikawa2017dirac}. Recent effort has pushed this ratio value to 13 \cite{wangke2018quantum}. All of the experimental work so far on graphene p-n junctions indicate that they fall far short of the ideal predictions. Theoretically, it is expected that atomic scale junction imperfections in real p-n junctions significantly modify their electron-optical functions \cite{libisch2017veselago}. These imperfections include junction interface roughness, finite junction width, and non-linearity as well as asymmetry of their doping profiles. It is thus important to measure the nanoscale properties of state-of-the-art p-n junctions, and to develop an understanding of how the non-ideality of a junction affects the transport of carriers across the interface. Scanning tunneling microscopy (STM) is an ideal probe to achieve this in graphene. It is a technique capable of giving atomic scale topographic and spectroscopic information across a single p-n junction, allowing us to characterize the junction completely \cite{lee2016imaging, gutierrez2016klein}. In this work, we use a four-probe STM with in-situ scanning electron microscopy (SEM) to study graphene p-n junctions. The four-probe STM allows for each individual probe to act independently as contacts for gating and bias as well as the scanning probe for STM/STS measurement. Our STM results are analyzed with Nonequilibrium Green's function (NEGF)-based simulation of electron flow through inhomogeneously doped graphene to develop a microscopic understanding of electron flow through realistic graphene p-n junctions.

\section{Results and Discussion}
    We perform our STM measurements on graphene p-n junctions that are fabricated by two independent techniques. Figure \ref{fig:fig1}b illustrates the first type of junction device based on a pre-patterned SiO$_2$/Si substrate with buried polysilicon gate electrodes, shown as red/orange bars in Fig. \ref{fig:fig1}b. These highly doped polysilicon gates are patterned to form interdigitated finger structures and buried within the SiO$_2$ layer. We study this junction as an example of what state-of-the-art CMOS-based processing is able to achieve for the junction geometry. All the red electrodes are electronically connected to a common contact pad, as are the orange electrodes. They then form coplanar split gates with 100 nm spacing within each pair. We create a stack of exfoliated graphene atop hexagonal boron nitride (hBN), and place it onto the buried gate SiO$_2$/Si substrate using the standard polymer dry transfer technique. A graphene p-n junction is formed above each pair of split gates. Also shown in Fig. \ref{fig:fig1}b is our experimental setup: two probes are used to apply the gate voltages to the buried polysilicon gates, a third probe (bias probe) is used to apply a sample bias to the graphene for STM measurement, and the final probe is used as the STM scanning probe. The second type of p-n junction is shown in Fig. \ref{fig:fig1}c. In this method, a piece of few-layered graphite with a naturally cleaved sharp edge is used as one of the gate electrodes, while the silicon wafer is used as the second gate. This method has been shown in transport experiments to be an improved choice for uniform, sharp gating for a new generation of graphene devices \cite{chen2016electron, zibrov2017tunable}. We use hBN to encapsulate the graphite gate. On top of the hBN/graphite/hBN stack sits a monolayer graphene flake, partially overlaying the underlying encapsulated graphite gate. In the experiment, independent back gate voltages are supplied to the bottom layer graphite and the underlying Si substrate to create a split gate on the monolayer graphene. Apart from the sharpness of the graphite gate, the graphite also avoids problems that arise from the grain structure of typical metals (patch effect). For the sake of discussion, we will refer to the first and second junction devices as the polySi and graphite gate devices respectively.
    \begin{figure*}[t]
    	\includegraphics[width=\linewidth]
    	    {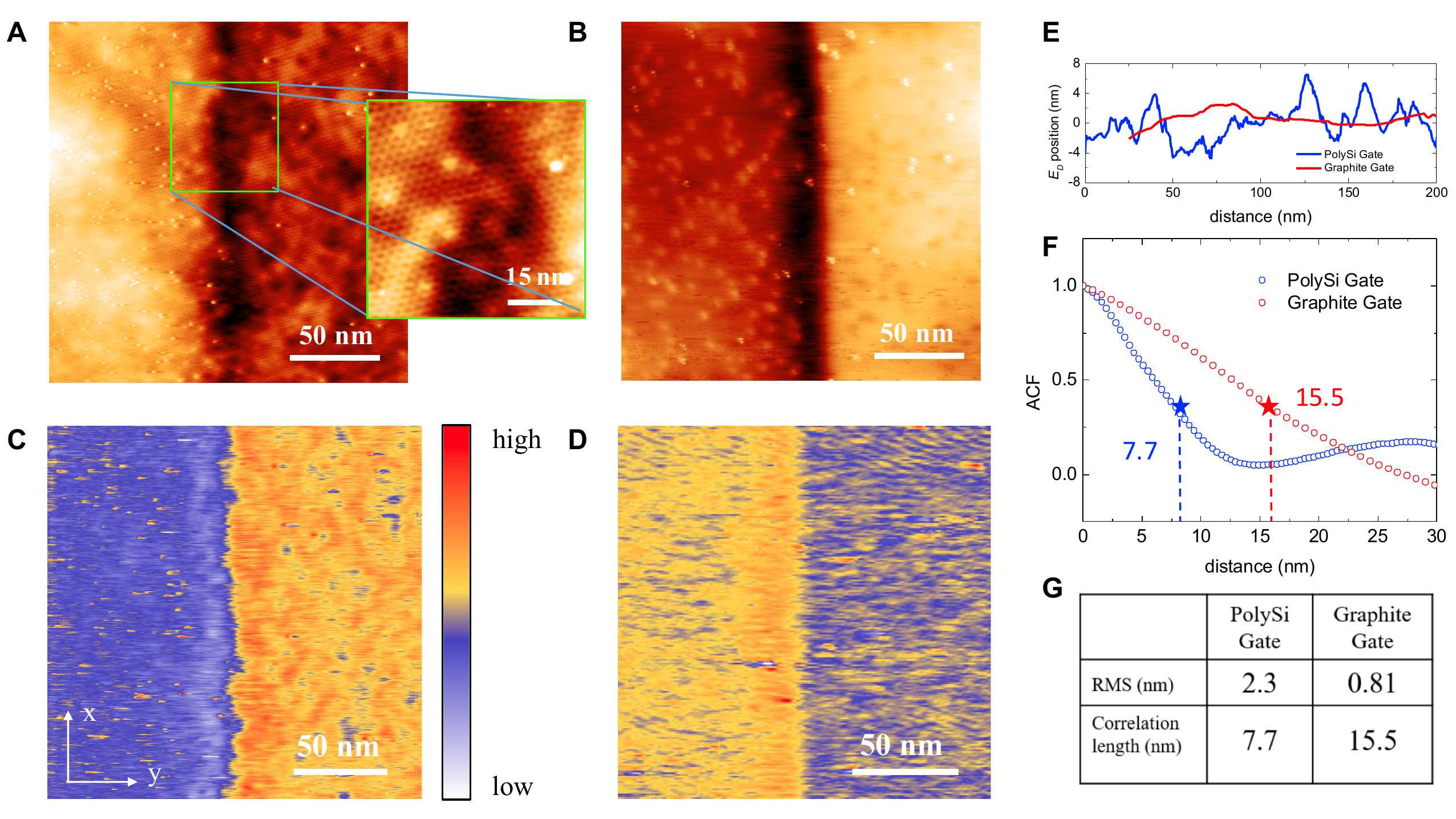}
    	\caption{
    		\label{fig:fig2}
    	    (\textbf{a}) A large STM topographic image of the junction interface in a polySi gate device. Inset is a magnified image of the junction area. The STM topography set points are -0.2 V, 30 pA and 0.3 V, 30 pA for the inset. (\textbf{b}) A large STM topographic image of the junction interface in a graphite gate device. STM topography set points are 0.1 V, 50 pA. (\textbf{c}) A $dI/dU$ mapping of the polySi gate device at the same area as Fig. \ref{fig:fig2}a. The image reflects a spatial distribution of Dirac point $E_D$ across the junction interface. Also labeled are $x$ and $y$ directions. (\textbf{d}) A $dI/dU$ mapping of the graphite gate device at the same area as Fig. \ref{fig:fig2}b. (\textbf{e}) Lateral interface roughness $y(x)$ extracted from $dI/dU$ mapping in Fig. \ref{fig:fig2}c-d. (\textbf{f}) Autocorrelation function of the profiles in Fig. \ref{fig:fig2}e for both devices. The correlation length  is marked by the star. (\textbf{g}) The RMS and correlation length from the analysis of Fig. \ref{fig:fig2}e.} 
        \end{figure*} 
    Our four-probe STM system is combined with an \textit{in-situ} UHV SEM, which allows us to accurately locate a sample area that only spans a few micrometers, and position the scanning probe precisely at the junction interface \cite{ji2012atomic}. Figures \ref{fig:fig1}d-e show SEM images of each device, showing contrast between differently gated regions allowing landing of the STM probe at the junction area as well as making electrical contacts for the gate and bias with the other probes. We note that with the in-situ four-probe capability in our system, no lithography is needed to pattern the electrodes. This greatly preserves the sample cleanliness, enabling high quality STM/STS measurements. Shown in Fig. \ref{fig:fig1}f is a large-scale STM topography of graphene on hBN taken on the polySi gate device showing a clear moir\'e pattern due to the mismatch between the graphene and hBN lattices. Imaging of large clean areas is essential to characterize the graphene p-n junctions, and this in turn is enabled by the four-probe instrument.
    \begin{figure*}[t]
    	\includegraphics[width=\linewidth]
    	    {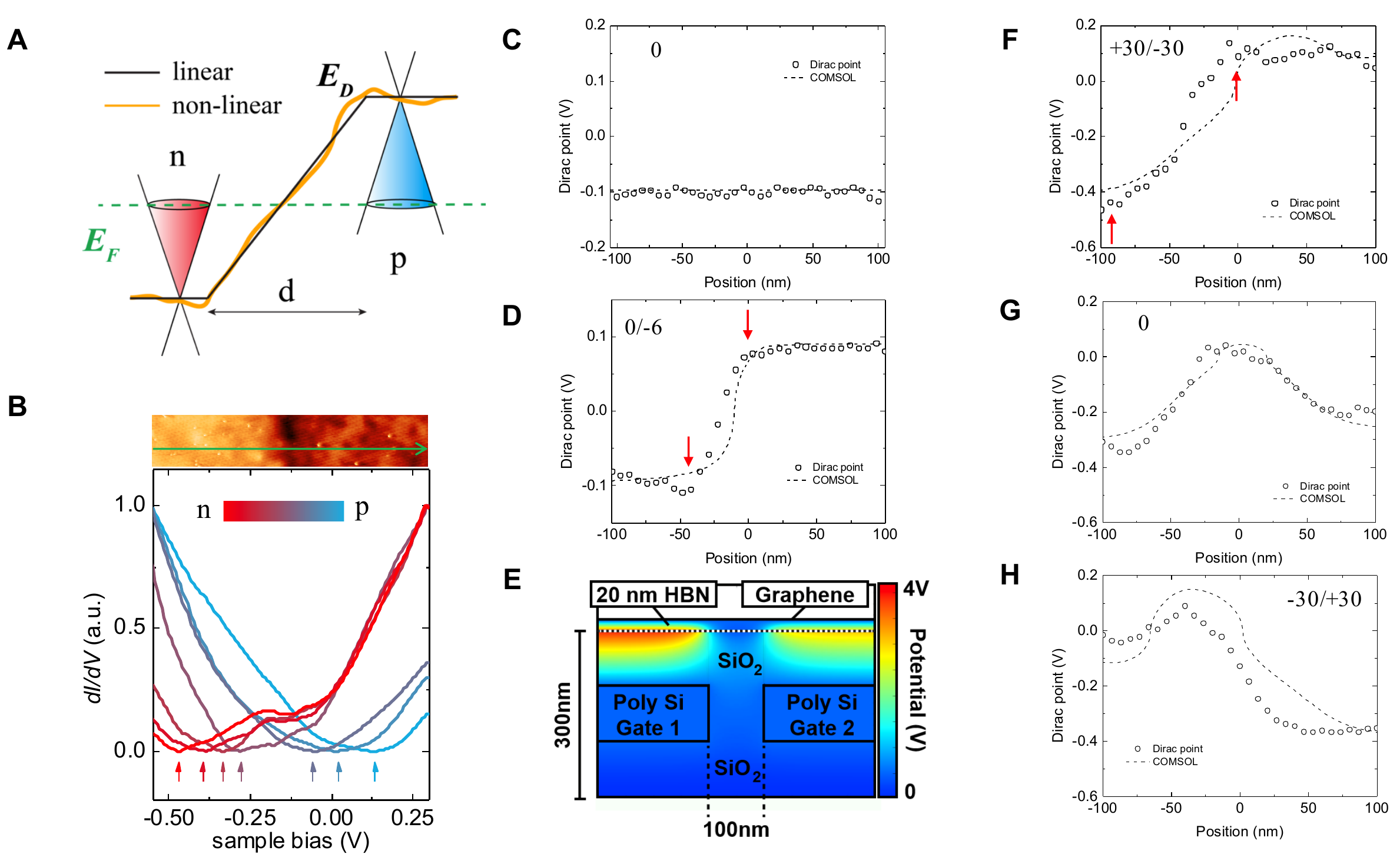}
    	\caption{
    		\label{fig:fig3}
    	    (\textbf{a}) Schematic of a theoretical potential profile of a graphene p-n junction with width $d$. The orange curve illustrates the non-linearity commonly existing in real junction devices. (\textbf{b}) Individual $dI/dV$ spectra selected from a line-cut taken along a green line (upper panel) across the junction interface in the polySi gate device. All spectra are normalized for display purpose. (\textbf{c-d}) Junction potential profiles of the graphite gate device determined from a line-cut spectra measurement under different gating conditions. COMSOL simulation results are also shown as dotted lines. (\textbf{e}) COMSOL simulation of built-in potential in the polySi gate device under zero gating. (\textbf{f-h}) Junction potential profiles of the polySi gate device under different gating conditions and corresponding COMSOL simulations.} 
        \end{figure*}
    
    Figure \ref{fig:fig2}a is a large STM topographic image across a p-n junction in a polySi gate device, with a gate bias of +30/-30 V applied to the two gates. Also shown is a zoomed-in image across a part of the junction where the moir\'e pattern between graphene and hBN is clearly seen. The junction itself is visualized as a step in topographic contrast across the image. We also show STM topographies of the same junction area at other gating conditions (see Supplementary Fig. \ref{fig:figS1}). This step feature at the junction is not visible in those conditions, indicating that the step seen in Fig. \ref{fig:fig2}a arises from electronic contrast in the density of states (DOS) across the junction rather than height contrast. The STM images in Fig. 2a and its inset also show several point defects. These defects are likely present in the hBN substrate \cite{wong2015characterization} and affect the graphene by electrostatic doping. To compare with the polySi gate geometry, Figure \ref{fig:fig2}b shows a STM topographic image of the graphite gate p-n junction with a gating condition of 0V/-6V. This image also shows the junction clearly as a step in topographic contrast. In the case of the graphite gate junction, a large part of the topographic contrast actually comes from the height change as the tip goes over the edge of the underlying graphite gate. Additional topographic contrast arises when the gates are biased to create a p-n junction (see Supplementary Fig. \ref{fig:figS2} for additional AFM and STM topography images on the graphite gate device).
    
    Simple visual inspection of Fig. \ref{fig:fig2}a and \ref{fig:fig2}b shows that real p-n junctions suffer from lateral roughness, and the graphite gate device has a significantly reduced roughness compared to the polySi gate device in this regard. To better characterize the lateral roughness of the junction, we performed spectroscopic $dI/dU$ mapping to measure the spatial local DOS (LDOS) at the energy of $E = eV_{bias}$ for both devices. $dI/dU$ mapping (a single $dI/dV$ value is taken at each pixel point) allows us to measure the LDOS with high spatial resolution in a short time compared to $dI/dV$ mapping (a whole $dI/dV$ spectrum is taken at each pixel point), thus minimize drift errors. Figure \ref{fig:fig2}c shows a $dI/dU$ map of the same region as Fig. \ref{fig:fig2}a, taken at $V_{bias}$ = -0.2 V. At this bias voltage, the DOS of p-doped graphene is higher than the DOS of n-doped graphene. Thus, the boundary between the two regions shows up clearly in the $dI/dU$ map with nanometer spatial resolution. We can then use this information to extract the lateral position $y$ of the boundary as a function of distance $x$ along the boundary. We do this in the following way: Graphene's DOS has a simple linear relationship DOS$(E) = A|E-E_D|$ where A is a constant and $E_D$ is the Dirac point energy. Therefore, a LDOS mapping at a fixed $E = eV_{bias}$ (such as Fig. \ref{fig:fig2}c) can be used to generate a mapping of $E_D$ in space with high resolution. $E_D$ will shift across the p-n junction ($y$ direction in Fig. \ref{fig:fig2}c). We pick the zero crossing of the Dirac point energy $E_D$ = 0 to define the spatial position of the boundary $y_{boundary}$. The variation of $y_{boundary}$ along the $x$ direction in Fig. \ref{fig:fig2}c thus characterizes the lateral roughness of the p-n junction. We plot such one-dimensional contour $y(x)$ as a blue line in Fig. \ref{fig:fig2}e. The same procedure can also be performed on the graphite gate device, as shown in Fig. \ref{fig:fig2}d and the red line in Fig. \ref{fig:fig2}e. We quantify the magnitude of the junction roughness by calculating a root mean square (RMS) value of the line profiles in Fig. \ref{fig:fig2}e. The RMS value of the roughness for the polySi gate device is 2.3 nm, compared to 0.81 nm for graphite gate. To analyze the lateral roughness of each junction in more detail, we calculate the autocorrelation function (ACF) of the junction profiles in Fig. \ref{fig:fig2}e, and show them in Fig. \ref{fig:fig2}f. We define a correlation length  as the point where the ACF has dropped to $1/e$ of its initial value. Figure \ref{fig:fig2}f shows that the roughness in the polySi gate has a shorter correlation length of $\lambda$ = 7.7 nm than that in the graphite gate where $\lambda$ = 15.5 nm. In Fig. \ref{fig:fig2}g, we compare the junctions numerically. The graphite gate device is smoother, having lower RMS and a larger correlation length for its junction roughness.
  
    Having characterized the junction roughness, we now turn to the actual doping profile as we transit from the hole doped to the electron doped side of the junction. In typical theoretical calculations of p-n junctions with width $d$, it is assumed that the doping profile linearly interpolates between the two sides of the junction as schematically illustrated in Fig. \ref{fig:fig3}a (black line). However, a real, non-ideal device inevitably possesses a certain non-linearity as denoted schematically by the orange line. To experimentally measure the junction potential profile, we take a series of $dI/dV$ spectra at equally spaced points along a line crossing the junction interface of the polySi gate device (green line at the top of Fig. \ref{fig:fig3}b). In Fig. \ref{fig:fig3}b, we show several selected $dI/dV$ spectra from a line-cut illustrating the evolution of the spectra across the junction. The STS spectrum of graphene has a minimum at the location of the Dirac point $E_D$. It is clearly seen that the Dirac point shifts from below to above the Fermi energy $E_F$ as the tip traverses the junction along the arrow direction, corresponding to a shift from \textit{n}-doped to \textit{p}-doped graphene. These line-cut $dI/dV$ profiles thus allow us to trace the spatial evolution of the Dirac point across the junction, and therefore the local doping profile of the graphene. In Figures \ref{fig:fig3}c-d we plot the spatial evolution of the Dirac point position $E_D$ relative to the Fermi level for the graphite gate junction for two separate gating conditions, 0V/0V and 0V/-6V. In the absence of gating, this region of graphene is \textit{n}-doped with the Dirac point $E_D$ at around -100 meV below $E_F$ (Fig. \ref{fig:fig3}c). Substrates are known to dope graphene in similar devices \cite{zhang2009origin}. When a voltage of -6V is applied to the graphite gate, a symmetric p-n junction is created as shown in Fig. \ref{fig:fig3}d. The situation is substantially more complicated in the case of the polySi gate devices. Shown in Fig. \ref{fig:fig3}f-h are the doping profiles for the polySi gate device for three separate gating conditions: +30V/-30V, 0V/0V and -30V/+30V, respectively. In Fig. \ref{fig:fig3}f, at +30V/-30V gate bias, two features of the profile stand out. Firstly, the measured potential profile displays a marked non-linearity as opposed to the ideal linear curve. Secondly, the measured junction has a very asymmetric potential profile, with higher doping on the \textit{n}-side than the \textit{p}-side, even under symmetric +30/-30V external gating conditions. We gain further insights into the device from the profile in Fig. \ref{fig:fig3}g at 0V/0V gate bias. Surprisingly, a pronounced doping distribution exists even without the gating. We attribute this to trapped charge from the fabrication process, as will be discussed below. There is a significant \textit{n}-doping on both sides of the junction with a slightly \textit{p}-doped region in the middle. Finally, Figure \ref{fig:fig3}h shows the doping profile under -30V/+30V gating. It is evident that we failed to create a p-n junction due to the inherent asymmetric doping in the device. We find similar results at different points along the junction. Our measurements of the local doping profiles in these graphene p-n junctions provide useful microscopic insight into the quality of the p-n junction, and confirm that lithographically fabricated gate structures are inferior to graphite based gate structures.  
    \begin{figure*}[t]
    	\includegraphics[width=\linewidth]
    	    {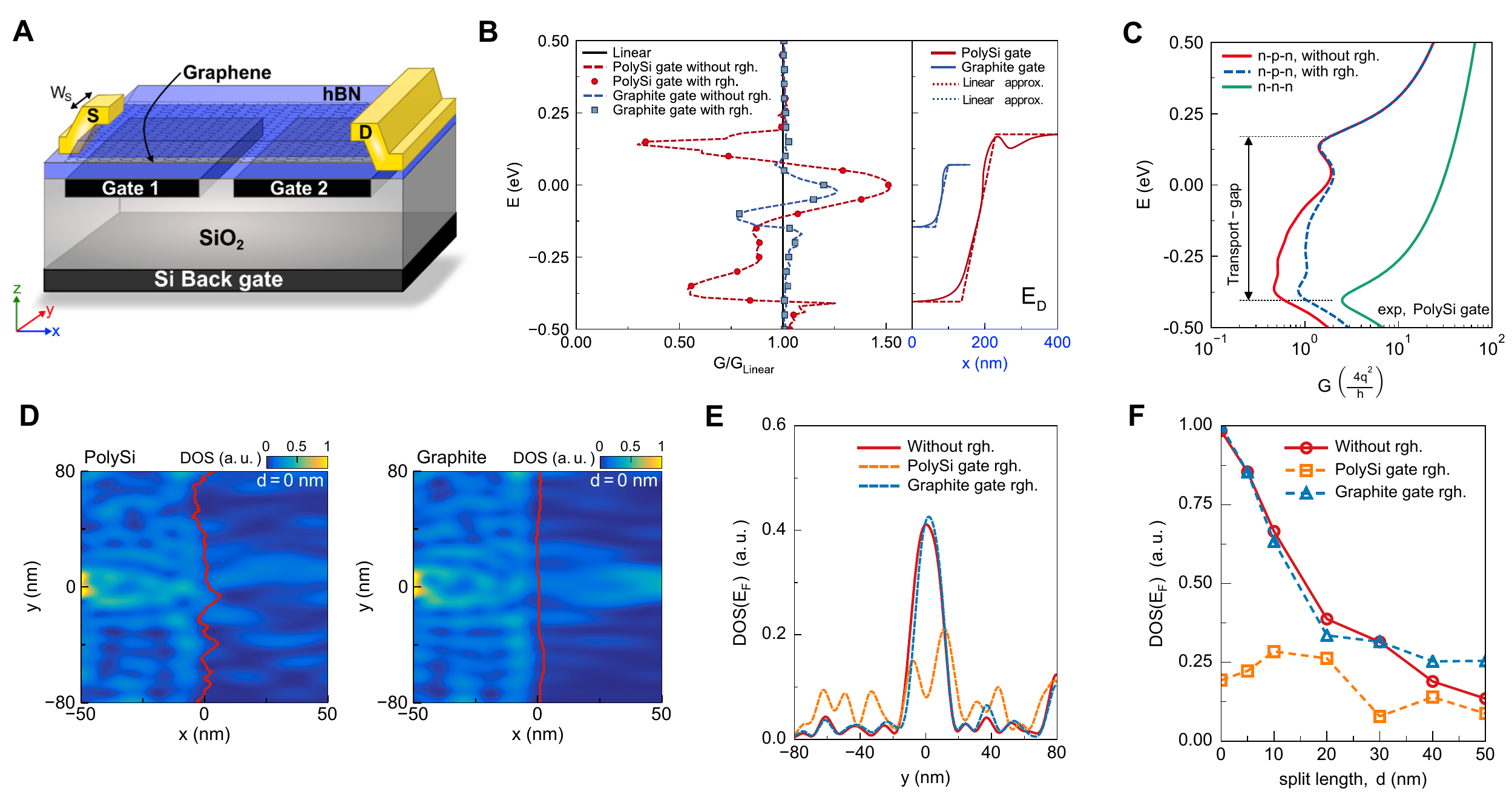}
    	\caption{
    		\label{fig:fig4}
    	    (\textbf{a}) Schematic of device structure to evaluate the performance of collimation and Veselago lensing under practical junction parameters. (\textbf{b}) Effect of non-linear potential in collimation of a single p-n junction. Junction interface roughness plays no role, whereas potential non-linearity causes the conductance to deviate from the linear case. (\textbf{c}) Effect of interface roughness on a double junction device. The added roughness increases the conductance resulting in a reduced current on/off ratio for such a device. (\textbf{d}) LDOS plot for the polySi and graphite gate devices. Both plots are for $d$ = 0 (abrupt junction) to show the effect of roughness alone. (\textbf{e}) Linecut at $x$ = 50 nm from Fig. \ref{fig:fig4}d. Focusing is evident for the ideal and graphite gate devices but not the polySi gate device. (\textbf{f}) Effect of junction width on focusing. Signal intensity at the focal point decays rapidly with increasing junction width due to reduced electron transmission.} 
        \end{figure*} 
    The presence of a doping distribution in the polySi gate junction indicates the presence of non-uniform charge under the graphene sheet even in the absence of gate voltages. We have attempted to quantitatively model this scenario using a simple electrostatic model where inhomogeneous trapped charge is located at the top of the SiO$_2$ surface. Supplement \ref{sec:S3} provides details of the simulation. We attribute the substrate doping to the fabrication process of the buried polysilicon gate electrodes, in particular the electro-polishing step used to smooth the SiO$_2$ above the polySi gates. In our simple model, we assume a uniform trapped charge density above each polySi gate electrode (independent of applied gate voltage), and use electrostatic COMSOL simulations of the real junction geometries, materials and potentials of our devices to best fit the measured doping profiles at all gating conditions. Figure \ref{fig:fig3}e shows the built-in potential in the polySi gate device calculated by our COMSOL model for zero gating condition with the trapped charge necessary to fit our three gating conditions as shown by the dotted lines in Fig. \ref{fig:fig3}f-h. To fit the three data sets optimally requires a piecewise trapped charge distribution above the SiO$_2$ of 7.5$\times$10$^{12}$ charges/cm$^2$ above the left gate, 5.6$\times$10$^{12}$ charges/cm$^2$ above the right gate and -7.5$\times$10$^{11}$ charges/cm$^2$ above the area between the gates with linear transitions between them. Although the confinement of the trapped charge to the surface is a simplistic approximation, the good agreement between models and doping profiles in Fig. \ref{fig:fig3}f-h suggests that trapped charge is a consistent explanation for the potential profiles seen at all three gating conditions. For the graphite gate, a uniform trapped charge density of 1.6$\times$10$^{12}$ charges/cm$^2$ above the SiO$_2$ creates the best fit of the model to the data. We also quantify the non-linearity of the potential profile for both devices by applying a linear fitting to the experimentally measured profiles and using the residual sum of squares as a figure of merit (see Supplementary Fig. \ref{fig:figS3}). The residual sum of squares from polySi gates is an order of magnitude larger than from graphite gates, implying a significantly larger non-linearity, as evident in Fig. \ref{fig:fig3}c-h. Although it is impossible to escape junction non-linearity in full due to the finite distance from the gating electrodes to the graphene in any device, the graphite gate device architecture is superior to the polySi as its graphite gating electrode is closer to the graphene and it lacks the variable trapped charge distribution. One can also extract the p-n junction width $d$ from these profiles. In Fig. \ref{fig:fig3}d, we use red arrows to denote the start and the end of the junction interface giving rise to a junction width $d$ of around 40 nm for graphite gates. Similarly, in Fig. \ref{fig:fig3}f, $E_D$ starts to change from the center and almost saturates approaching the left edge (red arrows). The junction width $d$ is thus around 100 nm for the polySi gate device, as expected since the spacing between the split gates was also 100 nm. Overall, the graphite gate device shows a narrower junction width, a smaller junction non-linearity and a more uniform substrate doping, allowing for gating to produce a symmetric profile, not possible in the polySi gate devices at reasonable voltages.\\ 
    
    Having characterized the parameters at the atomic scale in real graphene p-n junctions, we now assess the impact of these parameters on two key electron-optical functions: collimation and Veselago lensing. We do this through model simulations using the NEGF formalism. The analysis is helped by the fact that the width of the junction  (40-100 nm) is much larger than the length scale of the lateral roughness (a few nm) in the junction. We can therefore independently model the effect of lateral roughness and of the doping profile in our junctions. Veselago lensing happens for a sharp p-n junction where $k_F d \ll 1$, while carrier collimation happens for $k_F d \gg 1$. For our graphite gate device with a symmetric profile (a single $k_F$ value for both electron and hole doped side), a rough estimate gives $k_F d \approx 1$. For the polySi gate device which has an asymmetric profile (different $k_F$ values for electron and hole doped sides), a rough estimate gives $k_F d \approx 1 \sim 10$. These estimates put our junction devices in an intermediate regime between ideal collimation and ideal Veselago lensing. Given these considerations, we proceed to study what one would observe for each case with the real junction parameters. Figure \ref{fig:fig4}a shows the device structure in our model where we have an electron injection source with finite contact width $W_s$ on the left side of the junction and an extended drain on the right side to reduce multiple scattering from the edge and quantum interference effects. This device is a generic structure for both polySi gate and graphite gate devices (in the graphite case, Gate 2 region is controlled by the back gate).
    
    We first look at collimation. We use an extended source ($W_S$ = device width (160 nm) for Fig. \ref{fig:fig4}b-c) to evaluate the performance as this permits electrons to arrive with various incidence angles allowing us to assess the angular filtering function. We calculate the low bias conductance G of a single p-n junction at 300 K,
    \begin{equation}
        G (E_F) = \frac{4q^2}{h} \int T(E) \left( -\frac{\partial f_0}{\partial E} \right) dE
    \end{equation}

    Here, $T(E)$ is the transmission probability, $f_0 = f(E-E_F)$ is the Fermi function, $q$ is the electron charge and $h$ is Planck's constant (see Supplementary \ref{sec:S5} for calculation details). $G$, the conductance, characterizes the efficiency of electron transmission through the p-n junction. A higher conductance thus indicates a lower collimation, and vice versa. We first calculate the conductance for a linear doping profile with no roughness, which we use as a baseline for our calculations of real junctions. We then consider both the graphite and the polySi gate structures, and calculate the transmission through the junction with and without lateral roughness included in the calculation. The right panel of Fig. \ref{fig:fig4}b shows the experimentally measured non-linear junction profile and its linear approximation for both devices which are both used in the calculation. Finally, the conductance for each case is scaled by the conductance for the linear doping profile, and the resultant conductance ratios are plotted in Fig. \ref{fig:fig4}b. We clearly see that conductance of both devices deviates from the linear case in the p-n junction regime (-0.4 eV $<E<$ 0.175 eV for polySi and -0.15 eV $<E<$ 0.07 eV for graphite). This deviation is highest for the energy values close to the energy boundary of the junction where the slopes of the non-linear potential also differ most compared to the linear approximation. The deviation from unity in the conductance ratio is due to a variation in the slope for the non-linear case: a smaller slope of the potential is associated with a larger junction width $d$, which, in turn, creates a longer effective barrier (from the transmission probability formula stated above) and hence lower transmission probability (and hence higher collimation) and lower conductance. Similarly, larger slopes lead to higher transmission/conductance and lower collimation. The disadvantage of having a non-linearity is that the conductance $G(E_F)$ (hence the collimation) now strongly depends on the actual energy position of the Fermi level $E_F$ which itself is hard to control in real junction devices due to the substrate doping. In other words, the non-linearity makes collimation performance unpredictable, and therefore is a disadvantage from a design point of view. At the level of a single junction, the non-linearity in the junction doping profile plays an important role in determining the conductance of the junction, while the roughness does not have a significant impact on the conductance. 
    
    Although roughness does not greatly influence single junction conductance, it does affect the conductance of a two junction device. Sajjad \textit{et al.} proposed using a graphene double p-n junction for electronic switching, taking advantage of a highly angle selective transmission \cite{sajjad2013manipulating}. For such a two junction device where electron transmission through each junction is coupled with the other, one can create an effective gate-tunable ``transport gap" in graphene to turn the current on and off by using it in a unipolar regime or bipolar regime (p-n junction). We have calculated the electron conduction through such a two junction device. Figure \ref{fig:fig4}c shows results for the polySi gate devices. The green curve corresponds to a unipolar regime while the red curve represents a bipolar regime with a much lower conductance within the ``transport gap". However, as we add the interface roughness, the conductance in the bipolar regime increases (blue dashed curve), implying a reduction in the effective current on/off ratio for such a device.

    Next we look at Veselago lensing. For the Veselago simulation, we adopt a point source in our model ($W_s$ (= 32nm) $\ll$ device width (=160 nm)) and a symmetric junction (equal doping on both sides) to evaluate the impact of the junction width and interface roughness on the focusing. We first consider an abrupt junction ($d$ = 0) to isolate the contribution of interface roughness. Figure \ref{fig:fig4}d shows the calculated LDOS for polySi gate and graphite gate devices. We also calculate the LDOS for the ideal junction without roughness as a comparison (result not shown in Fig. \ref{fig:fig4}d). The LDOS was calculated using LDOS$(x_i, y_i; E)=A(x_i, y_i; E)/2 \pi$ where $(x_i, y_i)$ is the coordinate of site $i$ and $A$ is the spectral function given by $A=2 \textrm{Im}\{G^R\}$ with $G^R$ being the retarded Green's function obtained from the recursive Green's function algorithm \cite{bruque2007electron}. LDOS represents the probability density profile in the channel. In the case of Veselago focusing, electrons coming from the left point contact (source) should converge to a point on the right side of the junction at an equal distance from the junction. Therefore, the probability of finding electrons in the vicinity of the focal point should be higher and hence LDOS should also be higher than in the rest of the channel. Figure \ref{fig:fig4}d suggests that Veselago focusing is robust against relatively low roughness of the graphite gate for an abrupt junction. However, focusing characteristic is strongly smeared out at the high roughness seen in the polySi gate, even for an abrupt junction. This is clearly shown in Fig. \ref{fig:fig4}e, where line-cuts of Fig. \ref{fig:fig4}d at $x$ = 50 nm are compared. For the polySi gate case, the junction roughness plays an important role in randomizing the electron paths as they transit across the junction in a manner similar to roughness on an optical lens, and this effect is clearly seen in the LDOS as well. As the roughness is reduced (as is the case with the graphite gate), the focusing characteristics of the junction reappear and the LDOS becomes indistinguishable from the ideal junction case. We also repeat this calculation by varying the junction width $d$ to see its impact on the focusing. In Fig. \ref{fig:fig4}f, we plot the central peak intensity (at $y$ = 0 nm) of the $x$ = 50 nm line-cut as a function of junction width $d$ for ideal, polySi gate and graphite gate roughness cases. Again, the graphite gate shows a comparable performance to the ideal case, while the polySi gate randomizes the electron trajectories so much that the signal at the center point fluctuates at all widths. When the junctions are smooth enough (ideal and graphite gate cases), the focusing decays rapidly as the junction width $d$ increases, since fewer electrons can penetrate the junction and thus the signal intensity is reduced significantly. Two lessons are learned from these simulations. Firstly, the graphite gate is uniform and smooth enough to realize Veselago lensing whereas the polySi gate is fundamentally limited in these respects. Secondly, even in a perfect junction, the junction width sets another limitation on lensing, thus preventing the graphite gate device from lensing even though it meets other requirements. In the future, one should consider exploring improved devices using graphite gates, for instance with thinner hBN, to utilize the uniformity and smoothness of the graphite gate junctions and to create sharper junctions where signatures of Veselago lensing may be seen in experiments.
\section{Conclusions}
    We have reported atomic level characterization of two state-of-the-art types of graphene p-n junction devices using STM/STS. We find inherent imperfections as each junction exhibits finite width and lateral roughness as well as a chemical potential profile non-linearity and asymmetry which are directly measured with STS. We show that a significant improvement is exhibited in these parameters if one adopts an exfoliated graphite gate device geometry. To investigate the impact of these imperfections, we use our experimental findings as inputs into graphene p-n junction simulations of two important electron-optical applications: collimation and Veselago lensing. For Veselago lensing, junction roughness makes it impossible for polySi gates to lens for any junction width; graphite gates, although smooth enough, are inhibited by a junction width too large to lens. Collimation, as characterized by the conductance, is robust against junction roughness, but significantly affected by the non-linearity in the potential profile. Our work represents a significant advance in characterizing and analyzing graphene p-n junction for both fundamental research and practical applications.

\section*{Acknowledgments}
We thank J. Tersoff for helpful discussions. This work was funded by the Semiconductor Research Corporation through the Institute for Nanoelectronics Discovery and Exploration (INDEX) center of the Nanoelectronics Research Initiative (NRI).
    
%


\widetext
\clearpage
\begin{center}
{\large Supporting Information}\vspace{5mm}\\
\textbf{\large Atomic scale characterization of graphene p-n junctions for electron-optical applications}
\end{center}
\begin{center}
\begin{small}
{\textbf{ Xiaodong Zhou,$^{1,5,7}$ Alexander Kerelsky,$^1$ Mirza M. Elahi,$^2$ Dennis Wang,$^1$ K. M. Masum Habib,$^{2,8}$ Redwan N. Sajjad,$^6$ Pratik Agnihotri,$^4$ Ji Ung Lee,$^4$ Avik W. Ghosh,$^{2,3}$ Frances M. Ross,$^{5,9,\dagger}$ and 
Abhay N. Pasupathy$^{1,\dagger}$}}\\ \vspace{2mm}
$^1$\textit{Department of Physics, Columbia University, New York, New York 10027, United States}\\ \vspace{0.2mm}
$^2$\textit{Department of Electrical and Computer Engineering, University of Virginia, Charlottesville, VA 22904, United States
}\\ \vspace{0.2mm}
$^3$\textit{Department of Physics, University of Virginia, Charlottesville, VA 22904, United States}\\ \vspace{0.2mm} 
$^4$\textit{College of Nanoscale Science and Engineering, The State University of New York at Albany, Albany, New York 12203, United States}\\ \vspace{0.2mm}
$^5$\textit{IBM T. J. Watson Research Center, Yorktown Height, New York 10598, United States}\\ \vspace{0.2mm} 
$^6$\textit{Microsystems Technology Laboratories, Massachusetts Institute of Technology, Cambridge, Massachusetts 02139, United States}\\ \vspace{0.2mm} 
$^7$\textit{Current address: Institute for Nanoelectronic Devices and Quantum Computing, Fudan University, Shanghai 200438, P.R.China}\\ \vspace{0.2mm} 
$^8$\textit{Current address: Intel Corp., Santa Clara, California 95054, United States}\\ \vspace{0.2mm} 
$^9$\textit{Current address: Department of Materials Science and Engineering, Massachusetts Institute of Technology, Cambridge, Massachusetts 02139, United States}\\ \vspace{0.2mm} 
$^{\dagger}$\textit{Correspondence to: \href{mailto:apn2018@columbia.edu}{apn2018@columbia.edu}; \href{mailto:fmross@mit.edu}{fmross@mit.edu}}\\ \vspace{0.2mm} 
\end{small}
\end{center}
\setcounter{equation}{0}
\setcounter{figure}{0}
\setcounter{table}{0}
\setcounter{page}{1}
\setcounter{section}{0}
\makeatletter
\renewcommand{\theequation}{S\arabic{equation}}
\renewcommand{\thefigure}{S\arabic{figure}}
\renewcommand{\thesection}{S\arabic{section}}
\section{STM topographic images of the polySi gate junction area at various bias/gating conditions}\label{sec:S1}
\begin{figure*}[h!]
    	\includegraphics[width=0.8\linewidth]
    	    {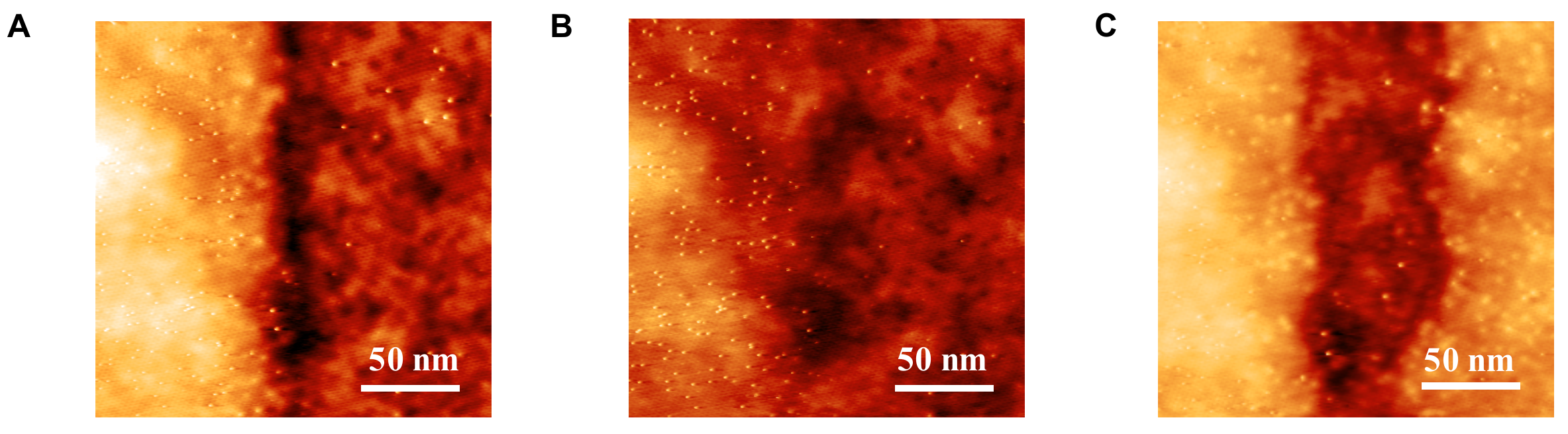}
    	\caption{\label{fig:figS1}
    	    (\textbf{a-c}) STM topographic images of a polySi gate device taken at different external gating conditions of (a) +30V/-30V, (b) -30V/+30V and (c) 0V/0V respectively.}
        \end{figure*}

    All STM topographic images were taken in a constant current mode. They show a convolution of electronic and structural contributions, but a feasible way to distinguish structural from electronic factors in STM topography is to collect the images at various sample bias conditions. Structural features should be independent of the sample bias while electronic features can change with bias. In our p-n junction devices, we can also change the external gating condition as this will change the density of states (DOS) landscape across the junction and therefore modulate the electronic contribution. In Fig. \ref{fig:figS1}a-c, we display three topographic images taken in the same junction area at different bias/gating conditions. Figure \ref{fig:figS1}a is the same as Fig. \ref{fig:fig2}a, taken at -0.2 V bias and +30 V/-30 V external gating condition. A `step edge' appears in the center of the image. Figure \ref{fig:figS1}b was taken at -0.5 V bias and -30 V/+30 V external gating. The `step edge' is less obvious in the image, indicating that it is not a true structural feature. In Fig \ref{fig:figS1}c, we turned off the external gating and collected the image at 0.1 V bias. Instead of a `step edge', a trench shows up in the image due to the change of DOS landscape across the junction. These images exclude any real structural step edge from the substrate as the origin of the `step edge' shown in Fig. \ref{fig:fig2}a.
    
\section{AFM/STM topographic images of the graphite gate junction area}\label{sec:S2}
\begin{figure*}[h!]
    	\includegraphics[width=0.9\linewidth]
    	    {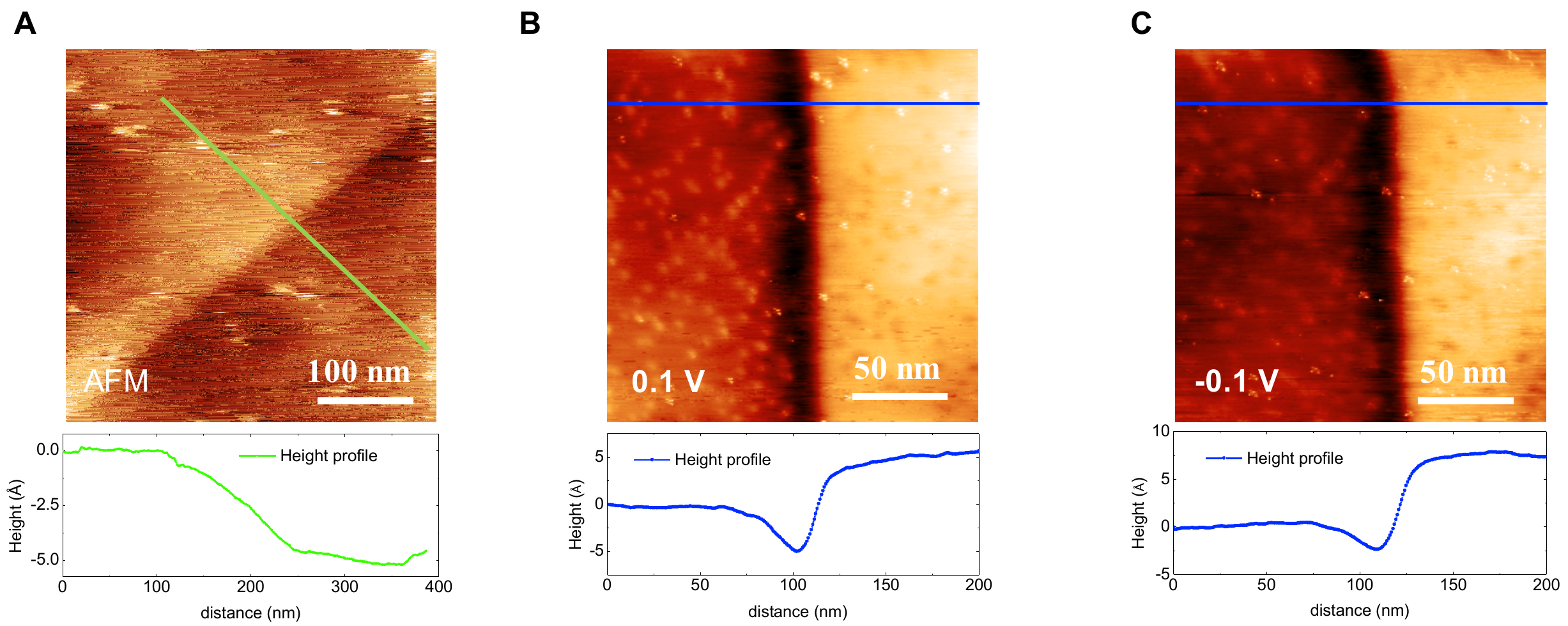}
    	\caption{\label{fig:figS2}
    	    (\textbf{a}) AFM image of the junction in a graphite gate device. (\textbf{b, c}) STM topographic image of the junction in a graphite gate device taken at 0.1 and -0.1 V bias, respectively, and 0V/-6V external gating.}
        \end{figure*}

    In a graphite gate device, the junction area should display a height contrast due to the underlying graphite gate. AFM across the junction, Fig. \ref{fig:figS2}a, displays a height contrast of 5 \r{A} at the junction, reflecting a true structural feature. Figure \ref{fig:figS2}b,c are STM topographic images taken at 0V/-6V external gating but with 0.1 V and -0.1 V bias, respectively. While the measured height change across the junction in Fig. \ref{fig:figS2}b is also around 5 \r{A}, consistent with the AFM result, it changes to 7.5 \r{A} in Fig. \ref{fig:figS2}c due to the relative change of local DOS at the two sides of the junction as we change the bias. We also note that the ``trench" seen at STM topography at the junction is also due to the reduced local DOS.
        
\section{Electrostatic calculation of the junction potential profile}\label{sec:S3}
    We have carried out an electrostatic numerical simulation of the junction potential profile in order to verify our hypothesis that trapped charge leads to experimental asymmetries and non-linearity. Simulations were done using finite-element modeling via the electrostatics module of the commercial software COMSOL. To model the polySi gate devices, we created a two-dimensional device geometry in COMSOL with piecewise junction geometry (Fig. \ref{fig:fig3}e in the main text) consisting of two polySi buried gates split by 100 nm, buried 100 nm below the surface of the 300 nm thick SiO$_2$ on silicon. Above the SiO$_2$ surface we modeled 20 nm of hexagonal boron nitride (hBN) (as in the experimental geometry). Next, the appropriate boundary conditions and voltages were applied to the gates, after which the potential profile was measured at the surface of the hBN, hence the graphene location. Finally, a graphene dispersion relation was applied to the potential profile to determine the piecewise Dirac point shifts. As this basic model clearly cannot reproduce the potential profiles with no gating nor the asymmetry for symmetric gating, we found that to reproduce experimental profiles, 3 different trapped charge densities at the surface of the SiO$_2$ are necessary: 7.5$\times$10$^{12}$ charges/cm$^2$ atop one polySi gate, 5.6$\times$10$^{12}$ charges/cm$^2$ above the second and -7.5$\times$10$^{11}$ charges/cm$^2$ between the two. These variable trapped charges likely arise due to the polishing involved to make the SiO$_2$ flat after burying the gates. A good agreement between the simulation with the given trapped charge values and the experiment is found, verifying that the asymmetry and non-linearity of our junction potential profile is consistent with these substrate-trapped charges.
    
\section{Analysis of non-linearity of junction potential profiles}\label{sec:S4}
\begin{figure*}[h!]
    	\includegraphics[width=0.9\linewidth]
    	    {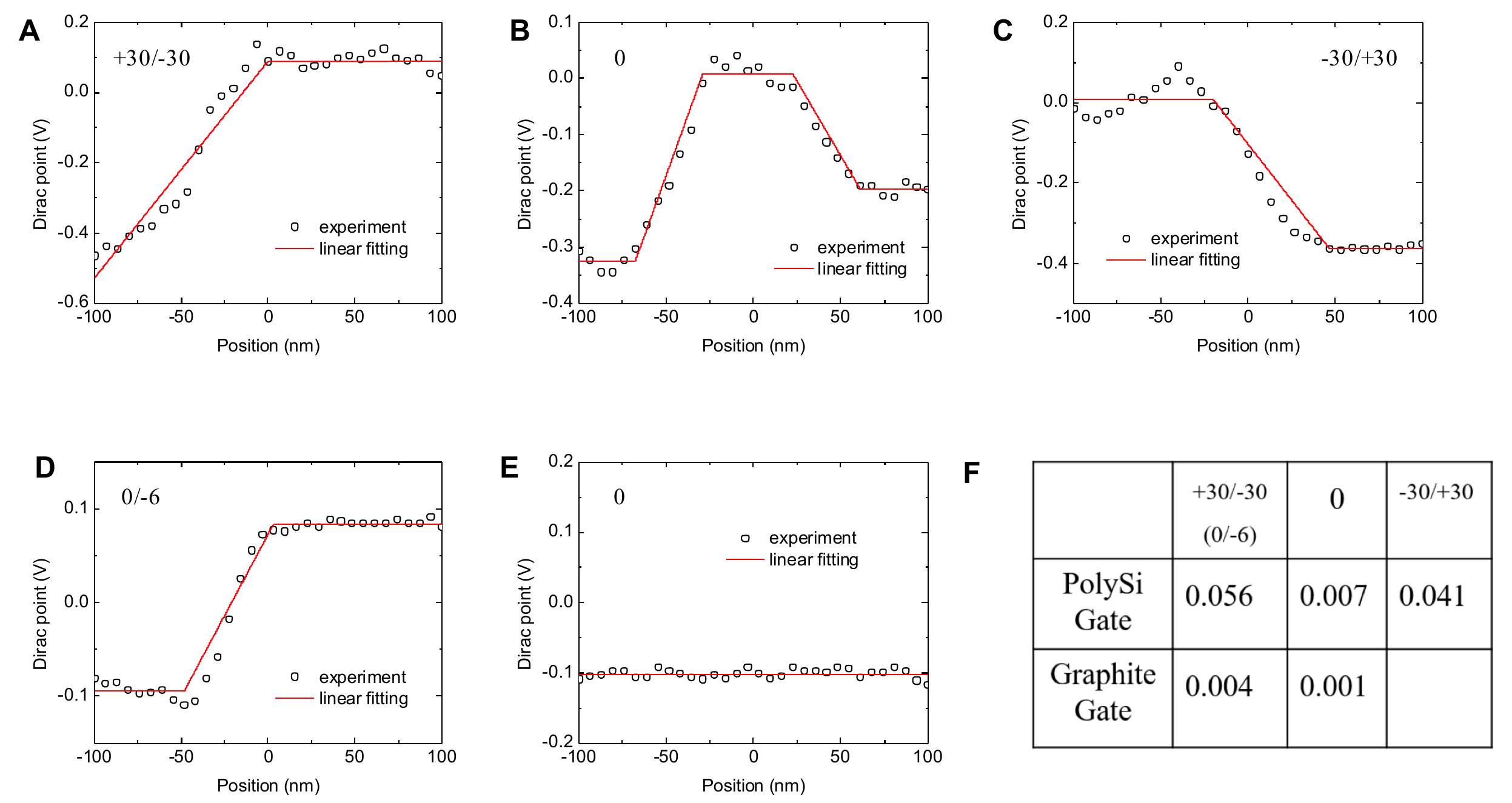}
    	\caption{\label{fig:figS3}
    	    (\textbf{a-e}) Experimental junction potential profiles of polySi gate and graphite gate devices and their respective linear fittings. (\textbf{f}) Table lists the residual sum of squares after fitting to quantify the non-linearity.}
        \end{figure*}

    To conduct a quantitative analysis of non-linearity of measured junction potential profiles, we apply linear fitting to the experimental profiles. Figure \ref{fig:figS3}a-e show the experimental curves and their linear fitting for both polySi gate and graphite gate devices. Then we calculate the residual sum of squares for each fitting and use it to quantify the non-linearity of junction potential profiles. Figure \ref{fig:figS3}f lists the results. The residual sum of squares for the polySi gate is an order of magnitude larger than that of the graphite gate meaning a larger non-linearity.
        
\section{Calculation of the conductance}\label{sec:S5}
    In a quantum transport simulation, the current is obtained using Landauer’s formula,
    \begin{equation}
        I = \frac{q}{h} \int T(E) \left[ f_1(E, \mu_1) - f_2(E, \mu_2) \right] dE
        \end{equation}
    where $T(E)$ is the transmission spectrum between contacts 1 and 2, $f(E, \mu_i)$ is the Fermi function of contact $i$, $q$ is the charge of an electron and $h$ is Planck's constant. For numerical efficiency, we have employed the recursive Green's function (RGF) algorithm [S1] where the channel is divided into $N$ blocks with block 1 connected to contact 1 and block $N$ connected to contact 2. In this approach, the transmission is calculated using a computationally efficient Green’s function formula,
    \begin{equation}
        T(E) = \textrm{tr} \{ \Gamma_{11} [  -2 \textrm{Im}(G_{11}) - G_{11} \Gamma_{11} G_{11}^\dagger ] \}
        \end{equation}    
    where $G_{11}$ is the Green's function for block 1 calculated using the RGF algorithm, and $\Gamma_{11} = -i (\Sigma_{11} - \Sigma_{11}^\dagger)$ with $\Sigma_{ii}$ being the self-energy of contact $i$ calculated using the decimation algorithm. The Hamiltonian matrix used in the above calculations is obtained using the modified  approach for computational efficiency [S2]. The electrostatic potential energy is modeled by modifying the diagonal elements of the Hamiltonian matrix by $-qV_i$ where $V_i$ is the electrostatic potential at side $i$.
    
\section*{References}

\noindent$^{S1}$  N. A. Bruque, M. Ashraf, G. J. Beran, T. R. Helander,and R. K. Lake, Physical Review B \textbf{80}, 155455 (2009).\\
$^{S2}$  K. M. Habib, R. N. Sajjad,   and A. W. Ghosh, AppliedPhysics Letters \textbf{108}, 113105 (2016).

\end{document}